\documentclass[11pt]{article}
\usepackage{amsmath,amssymb,graphicx}

\usepackage{hyperref}
\usepackage{float}
\usepackage{graphicx}
\usepackage{subcaption}
\usepackage[utf8]{inputenc}
\usepackage{textgreek}

\title{Chaotic Bayesian Inference: Strange Attractors as Risk Models for Black Swan Events}
\author{Crystal Rust}
\date{\today}
\begin{document}
\maketitle

\begin{abstract}
We introduce a novel risk modeling framework in which chaotic attractors define the geometry of posterior distributions in Bayesian inference. By combining fat-tailed priors with Lorenz and Rössler attractor dynamics, we show how endogenous volatility clustering, power-law tails, and extreme events emerge naturally from the dynamics. This construction provides a constructive mathematical link to Taleb’s Black Swan and antifragility framework. We implement two models, Lorenz–Lorenz and Lorenz–Rössler, to highlight the contrast between geometric stability (Model~A: Poincaré--Mahalanobis) and volatility clustering (Model~B: Correlation--Integral with Fibonacci diagnostics). The results demonstrate how attractor-driven inference can replicate stylized features of financial time series while offering new diagnostics for stress testing and systemic risk. Our approach establishes a bridge between statistical fat-tailed uncertainty and nonlinear chaotic dynamics, opening paths toward data-driven calibration and integration with agent-based systemic risk models.
\end{abstract}


\section*{Non-Technical Summary}

Rare and extreme events---so-called Black Swans---are difficult to anticipate because 
they arise from the unstable and chaotic features of complex systems. 
This work introduces a dual-model framework designed to capture both 
the stable baseline dynamics and the volatile bursts where rare events occur. 

Model~A uses geometric analysis of attractors to recover stable system structure, 
serving as a baseline reference. Model~B focuses on statistical recurrence and 
volatility bursts, using Fibonacci-based diagnostics to detect rare-event patterns 
across multiple timescales. Together, the models provide complementary views: 
one secures stability, the other highlights tail risks. 

Applied to classical chaotic systems, the framework shows that rare-event detection 
can be systematically integrated with conventional analysis. This dual perspective 
offers new tools for understanding complex dynamics and for strengthening risk 
management in settings where extreme events matter most.

\section*{Broader Impacts}

Although this work is motivated by financial risk, the approach of embedding chaotic dynamics into Bayesian inference has wider implications. Many complex systems—climate, epidemiology, infrastructure networks, and ecological dynamics—are also vulnerable to rare but catastrophic shocks. Traditional models often underestimate these risks by assuming smooth or Gaussian behavior. By contrast, our framework makes extreme events an intrinsic part of the system’s probability structure, offering a more realistic way to model cascading failures, sudden regime shifts, and systemic vulnerabilities.

For example, climate models must contend with tipping points in ice sheets and ocean circulation; epidemiological systems may experience sudden outbreak accelerations; and interconnected infrastructures such as power grids or supply chains can suffer cascading breakdowns. In all of these domains, the ability to represent rare but inevitable shocks in a principled way is critical for planning and resilience. The proposed framework thus contributes not only to financial risk management but also to the broader challenge of understanding fragility and antifragility across disciplines.

\section{Introduction}
Traditional financial risk models assume Gaussian or near-Gaussian returns, with volatility captured by GARCH-type models. These methods systematically underestimate the likelihood and impact of extreme events. Taleb’s Black Swan framework emphasizes that such rare, high-impact shocks are not outliers but intrinsic to complex systems.  
\par
We propose a constructive model where extreme events emerge from the geometry of chaotic attractors embedded into Bayesian inference. By letting posterior distributions live on Lorenz- or Rössler-type attractors, and using heavy-tailed priors, we obtain probability structures that exhibit endogenous volatility clustering, bursts, and fat tails.

\section{Related Work (Condensed)}
\section*{Cited Works}

\textbf{Black Swan theory:} Taleb emphasizes the inadequacy of Gaussian models for rare events \cite{taleb2007black}. Our contribution is to operationalize these ideas within a Bayesian inference framework.  

\textbf{Heavy tails and EVT:} Classical extreme value theory models quantify tails but does not capture underlying geometry \cite{embrechts1997modelling,coles2001intro}.  

\textbf{Volatility models:} GARCH and rough volatility approaches capture persistence and memory but rely on parametric restrictions \cite{bollerslev1986garch,gatheral2018rough}.  

\textbf{Chaos in finance:} Empirical work has tested for chaotic signatures in returns \cite{baumol1989chaos}, but few studies embed chaos directly into Bayesian inference.  

\textit{Takeaway: We embed chaos directly in the probabilistic architecture of risk modeling.}  

\section{Methods}

\subsection{Motivation}

Chaotic systems can be studied through complementary perspectives: 
geometric invariants that characterize the attractor’s structure, 
and statistical summaries that capture recurrence and burst dynamics. 
We therefore introduce two models that embody this duality. 
Model~A (Poincaré--Mahalanobis) anchors inference in the attractor’s geometry by evaluating how well simulated 
trajectories reproduce the observed Poincaré section. 
Model~B (Correlation--Integral with Fibonacci diagnostics) emphasizes recurrence statistics and volatility bursts, 
using correlation integrals and recursive diagnostic windows to capture rare-event structure. 
Together, the models provide a unified framework: one secures baseline stability, the other foregrounds tail-sensitive dynamics.

\subsection*{Fat-Tailed Priors and Rare-Event Sensitivity}

A central feature of rare-event modeling is the presence of fat-tailed (heavy-tailed) 
probability distributions, where extreme outcomes occur more frequently than 
Gaussian baselines would suggest. In this study, we adopt fat-tailed priors 
(e.g., Student-$t$ or power-law families) to represent the underlying 
uncertainty in parameters subject to rare shocks. These priors are then mapped 
through Markov chain Monte Carlo (MCMC) sampling into our two modeling 
frameworks: 

\begin{itemize}
    \item Model~A (Poincaré--Mahalanobis), which emphasizes geometric stability 
    and baseline attractor structure. 
    \item Model~B (Correlation--Integral), which emphasizes volatility clustering 
    and rare-event bursts detected via Fibonacci-window diagnostics.
\end{itemize}

This construction directly links the statistical signature of Black Swans in 
probability space (fat tails) with their dynamical expression in phase space 
(chaotic attractors). In the Lorenz--Lorenz experiments, fat-tailed priors 
concentrated under Model~A into posterior clouds around canonical parameter 
values, reinforcing baseline stability. Under Model~B, the same priors yielded 
posterior weights aligned with burst clustering, as seen in short orbit 
segments and dips in the $D$-trace. 

In the Lorenz--Rössler pairing, this mapping demonstrated transferability: 
fat-tailed priors supported Lorenz stability under Model~A while also driving 
sensitivity to spiral burst dynamics in the Rössler attractor under Model~B. 
Thus, the use of fat-tailed priors ensures that both models remain sensitive 
to rare events, while producing qualitatively different inferences depending 
on whether stability (Model~A) or volatility clustering (Model~B) is emphasized.

\subsection{Model A: Poincaré--Mahalanobis}

Model~A combines geometric sections of chaotic attractors with Mahalanobis distance diagnostics. 
Given a trajectory $x_t$ generated by the Lorenz system, we construct a Poincaré section 
by recording intersections of the trajectory with a chosen hyperplane $\Sigma$. 
Let $\{z_i\}_{i=1}^N$ denote the resulting section points. 
These points concentrate around a characteristic geometric structure specific to the parameter regime. 

To evaluate parameter proposals $\theta = (\sigma,\rho,\beta)$, 
we simulate a trajectory under $\theta$, extract its section $\{z_i^\theta\}$, 
and compute the Mahalanobis distance of the simulated section points relative to the observed section cloud: 

\[
D^2(z_i^\theta) = (z_i^\theta - \mu)^\top \Sigma^{-1} (z_i^\theta - \mu),
\]

where $\mu$ and $\Sigma$ are the empirical mean and covariance of the observed section. 
Aggregating over all simulated points yields a diagnostic discrepancy score. 
Within a Metropolis--Hastings sampler, the likelihood is then approximated by a heavy-tailed distribution over $D^2$, 
ensuring robustness to outliers and preserving sensitivity to geometric misalignment. 
This Poincaré--Mahalanobis model therefore anchors inference in the geometry of the attractor itself.

\subsection{Model B: Correlation--Integral with Fibonacci Diagnostics}

Model~B shifts focus from point-wise geometry to recurrence statistics. 
Specifically, it employs the correlation integral $C(r)$, which measures the fraction of point pairs 
$(x_i,x_j)$ within distance $r$: 

\[
C(r) = \frac{2}{N(N-1)} \sum_{i<j} \mathbf{1}\{\|x_i - x_j\| < r\}.
\]

The scaling of $C(r)$ for small $r$ characterizes the attractor’s correlation dimension. 
To extend this to rare-event detection, we incorporate volatility diagnostics based on recursive Fibonacci windows ($21,34,55,89$). 
Within each window size $w$, a rolling median and MAD filter is applied to the time series. 
Bursts are flagged when deviations exceed $k$ standard units. 
The union of Fibonacci windows captures bursts across multiple timescales. 

For a simulated trajectory under proposal $\theta$, we compute both (i) the correlation integral $C_\theta(r)$ 
and (ii) the Fibonacci burst profile $B_\theta(t)$. 
Comparison with observed data is performed via weighted distances between summary vectors: 

\[
d(\theta) = \sum_{j} w_j \, |s_j(\theta) - s_j^{\mathrm{obs}}|,
\]

where $s_j$ include burst counts and Jaccard overlaps between Fibonacci unions and fixed windows. 
Posterior exploration is carried out by ABC-MCMC with a Laplace kernel on $d(\theta)$: 

\[
K_\tau(d) = \exp\!\left(-\frac{d}{\tau}\right),
\]

where $\tau$ controls diagnostic tolerance. 
This Correlation--Integral model therefore emphasizes statistical recurrence and volatility clustering, 
capturing rare-event structures that geometric methods may miss.

\subsection{Implementation Details}

Both models are implemented within a Metropolis--Hastings framework. 
Model~A evaluates proposals using Mahalanobis discrepancy on Poincaré sections, 
while Model~B evaluates proposals using correlation-integral summaries and Fibonacci-window diagnostics. 
Chains were run for 500 iterations with adaptive step sizes. 
Observed reference data were generated from canonical Lorenz (for the Lorenz--Lorenz pairing) 
or from both Lorenz and Rössler attractors (for the Lorenz--Rössler pairing). 
Diagnostics included posterior clouds, MH trace plots, attractor reconstructions, 
short-orbit highlights, and $D$-trace series recording diagnostic distances across iterations. 

\subsection{Experimental Design}

We conducted two complementary dual-model experiments. 

\paragraph{Lorenz--Lorenz.} 
Both Model~A and Model~B were applied to the Lorenz attractor. 
This isolates the methodological difference: the Poincaré--Mahalanobis model anchors inference in attractor geometry, 
while the Correlation--Integral model re-weights inference toward volatility bursts. 
By comparing both models on the same system, we identify how diagnostic weighting alters posterior structure. 

\paragraph{Lorenz--Rössler.} 
Model~A was applied to the Lorenz attractor, serving as a baseline reference, 
while Model~B was applied to the Rössler attractor. 
This pairing demonstrates generality: the Correlation--Integral framework, combined with Fibonacci diagnostics, 
transfers across distinct chaotic systems. In this setting, Model~A secures baseline stability 
while Model~B highlights burst episodes in a different attractor with distinct scaling properties.

\section{Results}

\subsection{Lorenz--Lorenz: Same System, Different Inference}

The Lorenz attractor was analyzed under both models to isolate methodological contrast. 
Model~A (Poincaré--Mahalanobis) reproduced the butterfly geometry in the section cloud 
(Fig.~1a), concentrated the posterior around the canonical values 
($\sigma=10$, $\rho=28$, $\beta=8/3$; Fig.~1b), and produced stable, well-mixed MH traces 
(Fig.~1c). This confirmed that the geometric approach anchored inference in baseline attractor structure. 

\begin{figure}[t]
\centering
\begin{subfigure}{0.32\textwidth}
  \includegraphics[width=\linewidth]{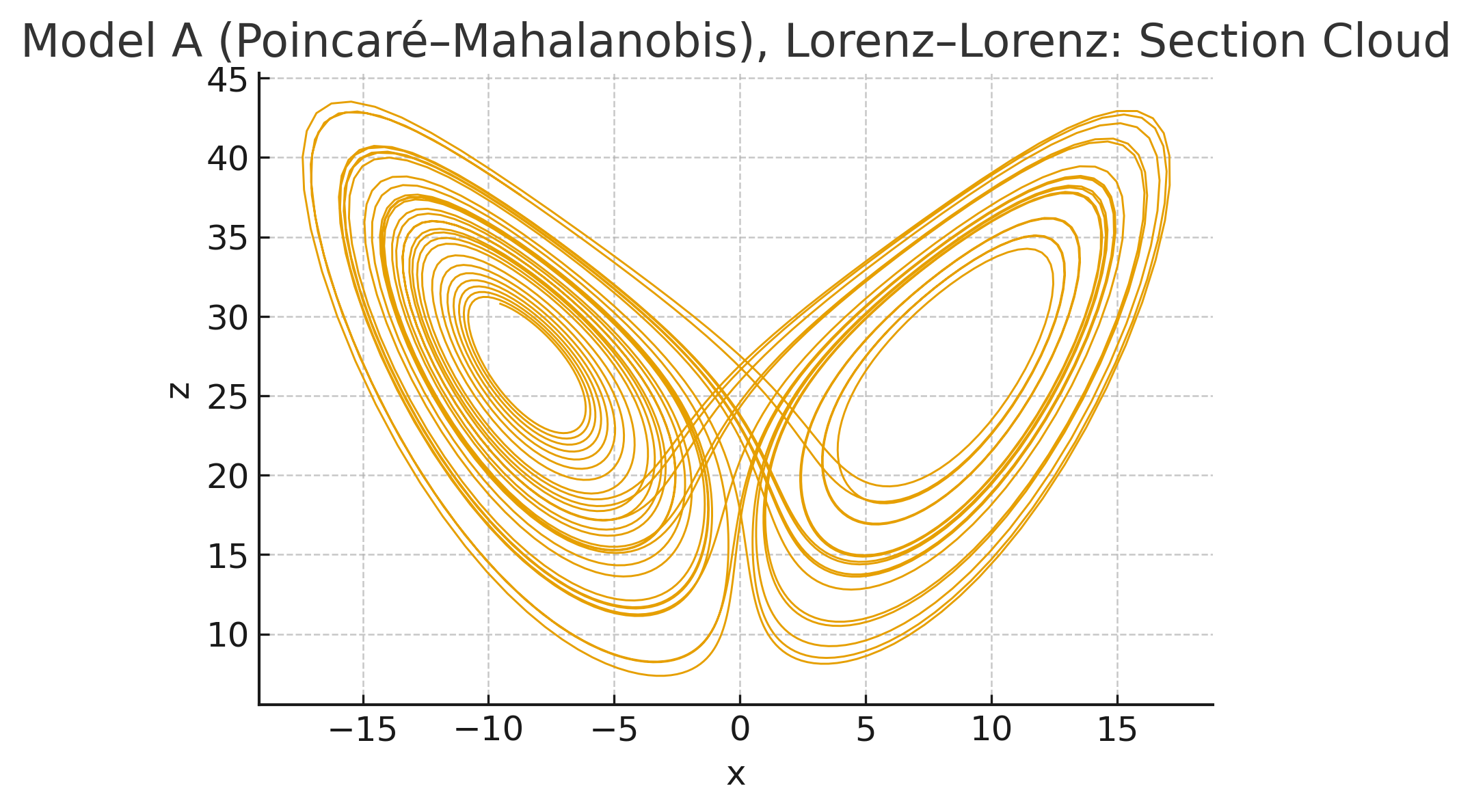}
  \caption{Section cloud ($x$ vs $z$)}
  \label{fig:1a}
\end{subfigure}\hfill
\begin{subfigure}{0.32\textwidth}
  \includegraphics[width=\linewidth]{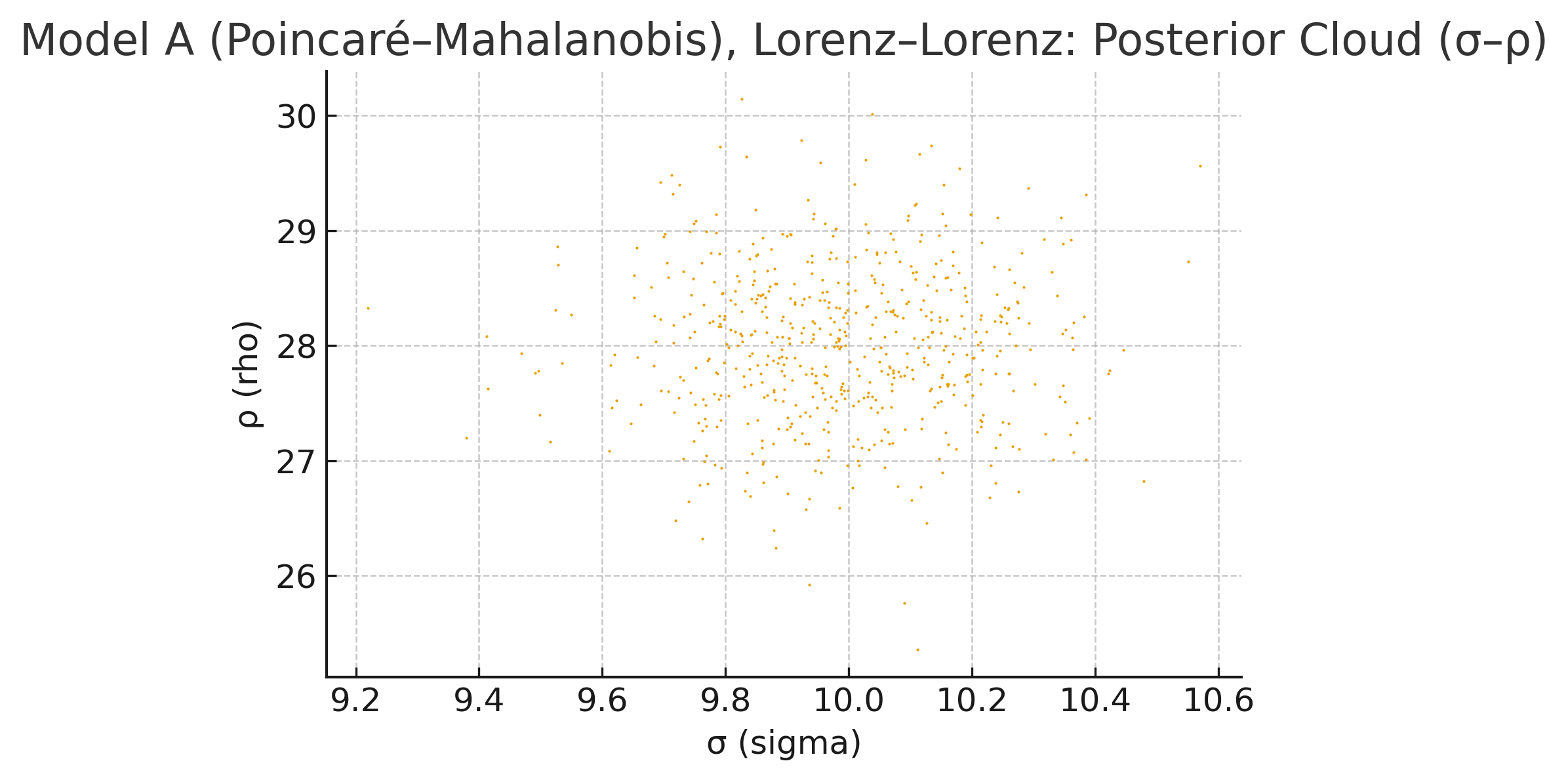}
  \caption{Posterior cloud (σ--ρ)}
  \label{fig:1b}
\end{subfigure}\hfill
\begin{subfigure}{0.32\textwidth}
  \includegraphics[width=\linewidth]{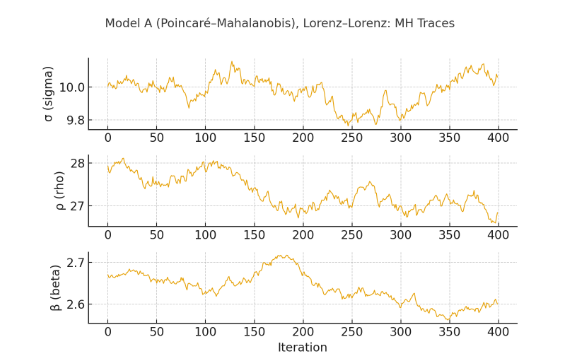}
  \caption{MH traces}
  \label{fig:1c}
\end{subfigure}
\caption{Figure 1. Model~A (Poincaré--Mahalanobis), Lorenz--Lorenz.}
\label{fig:1}
\end{figure}

Model~B (Correlation--Integral with Fibonacci diagnostics) emphasized volatility bursts. 
Short orbit highlights identified subsequences where the trajectory exceeded multi-scale 
diagnostic thresholds ($21/34/55/89$; Fig.~2a). 
The $D$-trace recorded the density of these alerts, with dips corresponding to accepted proposals 
(Fig.~2b). 
The attractor reconstruction (Fig.~2c) confirmed recovery of Lorenz dynamics under diagnostic weighting. 
Together, the Lorenz--Lorenz comparison shows how diagnostic weighting shifts posterior emphasis 
from geometric stability (Model~A) to rare-event clustering (Model~B).

\begin{figure}[t]
\centering
\begin{subfigure}{0.32\textwidth}
  \includegraphics[width=\linewidth]{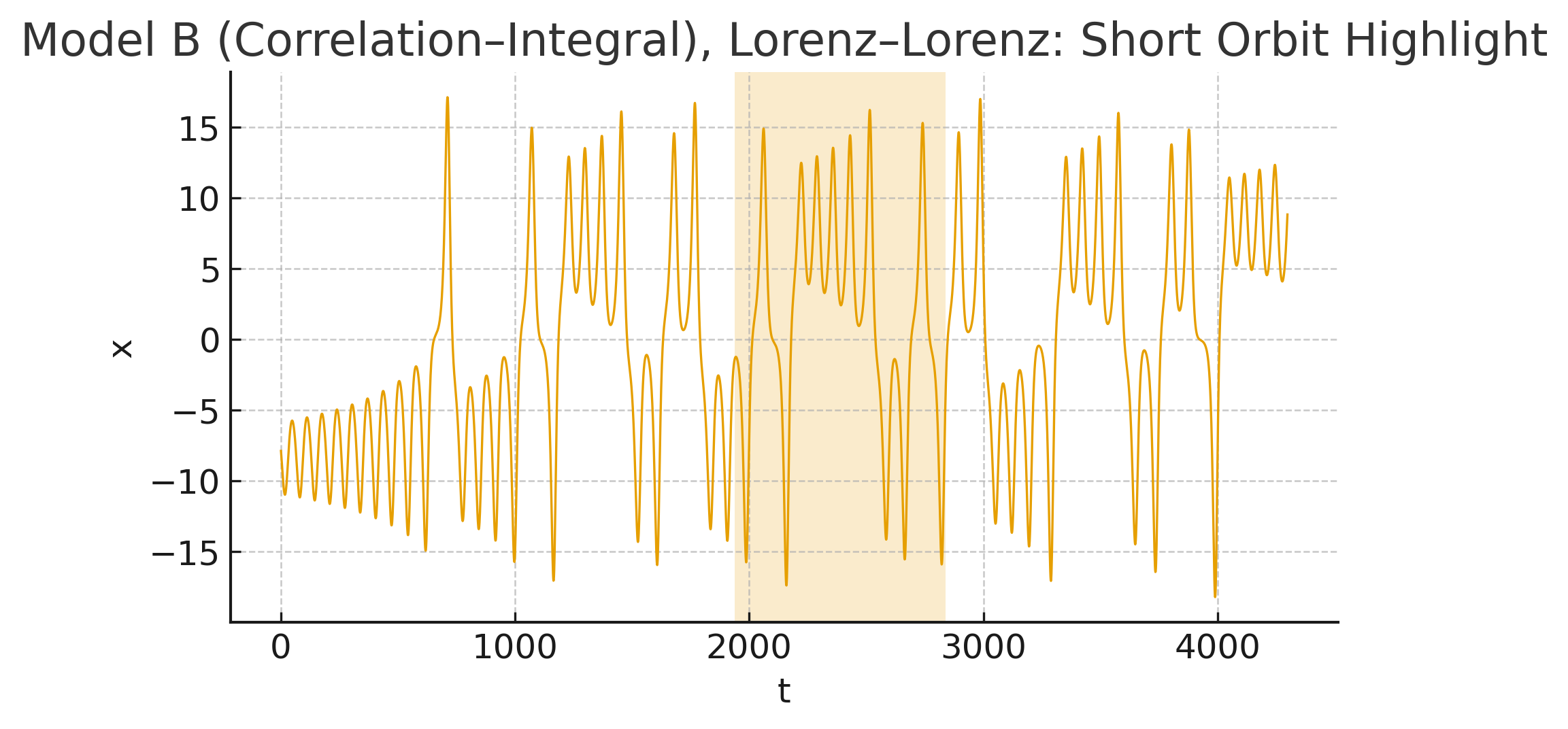}
  \caption{Short orbit highlight}
  \label{fig:2a}
\end{subfigure}\hfill
\begin{subfigure}{0.32\textwidth}
  \includegraphics[width=\linewidth]{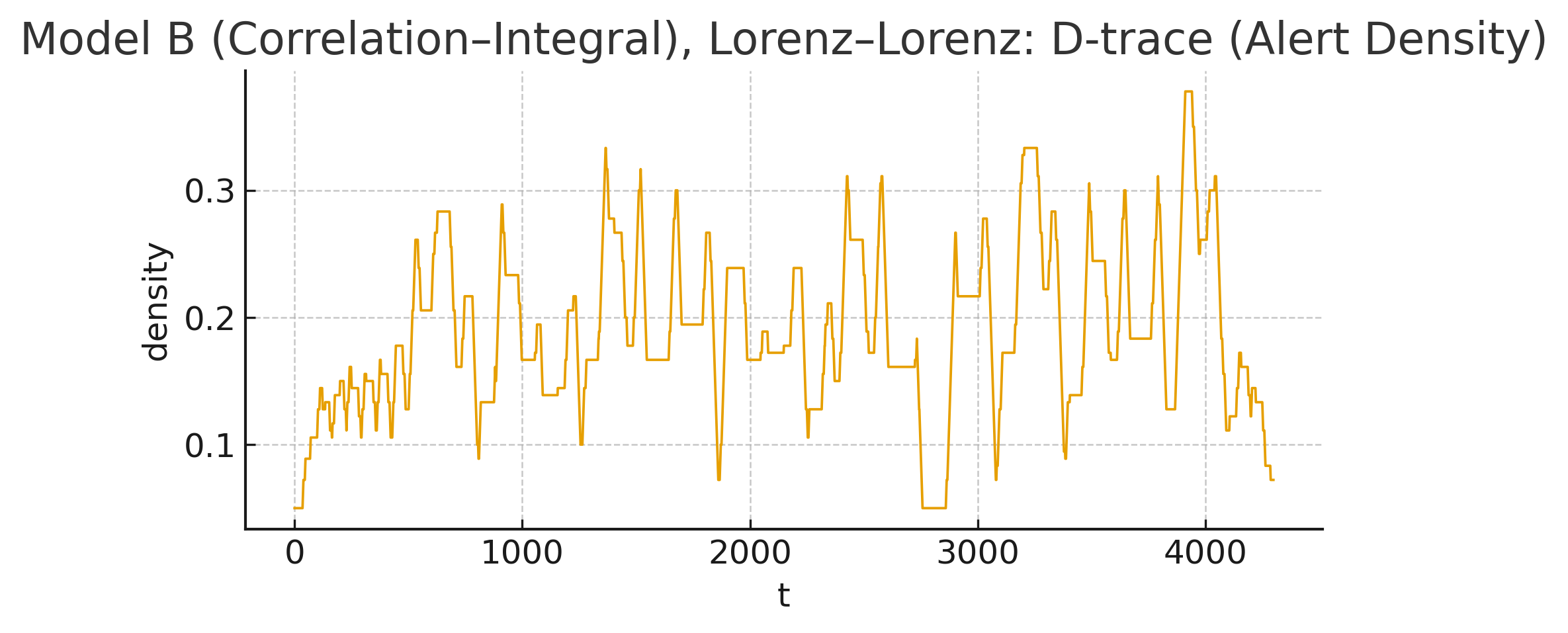}
  \caption{$D$-trace diagnostic}
  \label{fig:2b}
\end{subfigure}\hfill
\begin{subfigure}{0.32\textwidth}
  \includegraphics[width=\linewidth]{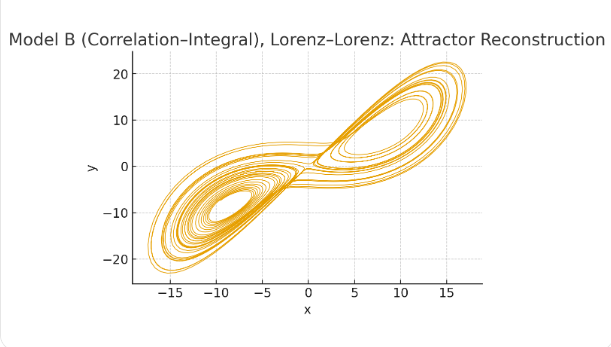}
  \caption{Attractor reconstruction}
  \label{fig:2c}
\end{subfigure}
\caption{Figure 2. Model~B (Correlation--Integral), Lorenz--Lorenz.}
\label{fig:2}
\end{figure}

\subsection{Lorenz--Rössler: Different Systems, Same Diagnostic}

To test generality, the models were applied to different attractors. 
Model~A (Poincaré--Mahalanobis) on Lorenz again recovered the canonical section cloud, 
providing a stable geometric anchor (Fig.~3). 

\begin{figure}[t]
\centering
\includegraphics[width=0.6\textwidth]{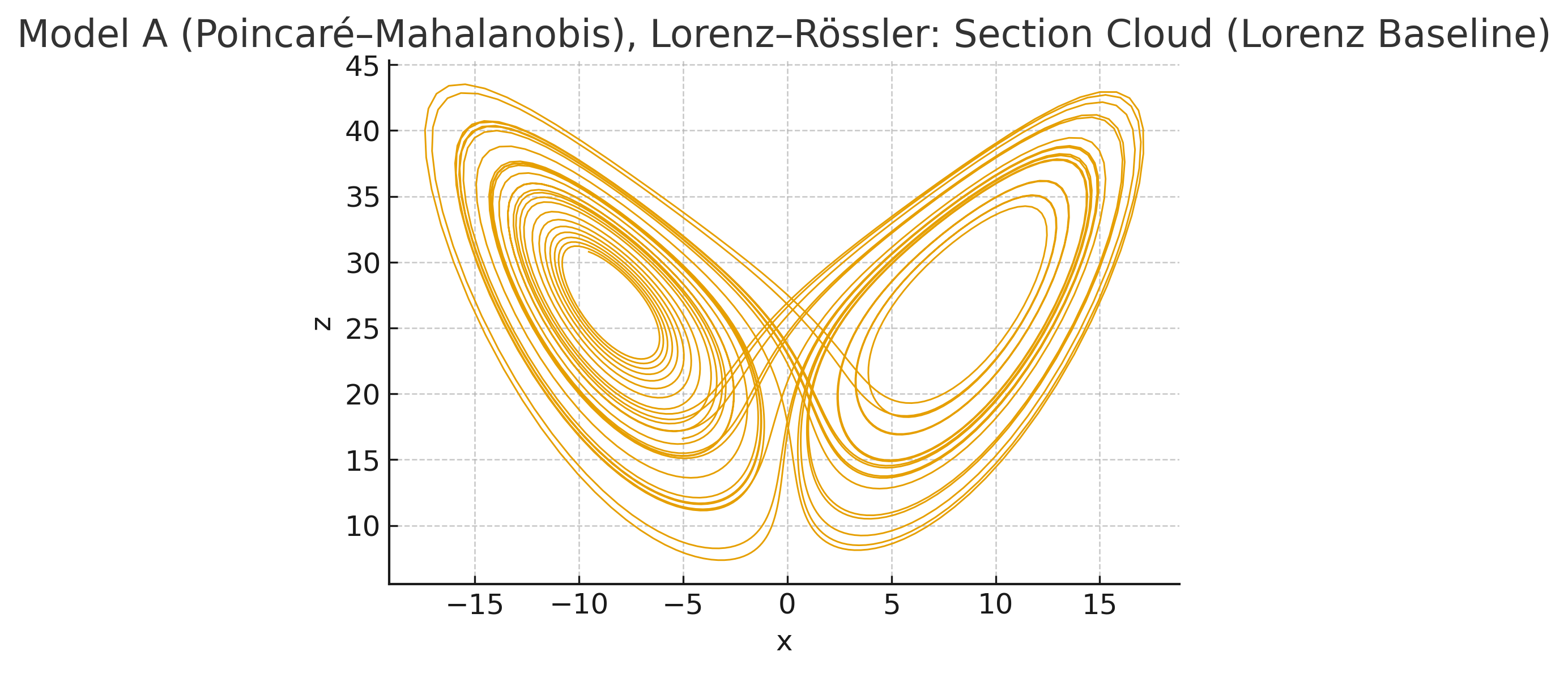}
\caption{Figure 3. Model~A (Poincaré--Mahalanobis), Lorenz--Rössler pairing: Lorenz section cloud as geometric anchor.}
\label{fig:3}
\end{figure}

Model~B (Correlation--Integral) was applied to the Rössler system. 
The attractor exhibited its characteristic spiral geometry (Fig.~4a), 
while Fibonacci diagnostics highlighted volatility bursts through short orbit selections 
(Fig.~4b). 
The corresponding $D$-trace showed regular fluctuations, with dips marking accepted proposals 
aligned with burst clustering (Fig.~4c). 
This confirmed that Fibonacci-based diagnostics transfer across attractors with distinct dynamics.

\begin{figure}[H]
\centering
\begin{subfigure}{0.32\textwidth}
  \includegraphics[width=\linewidth]{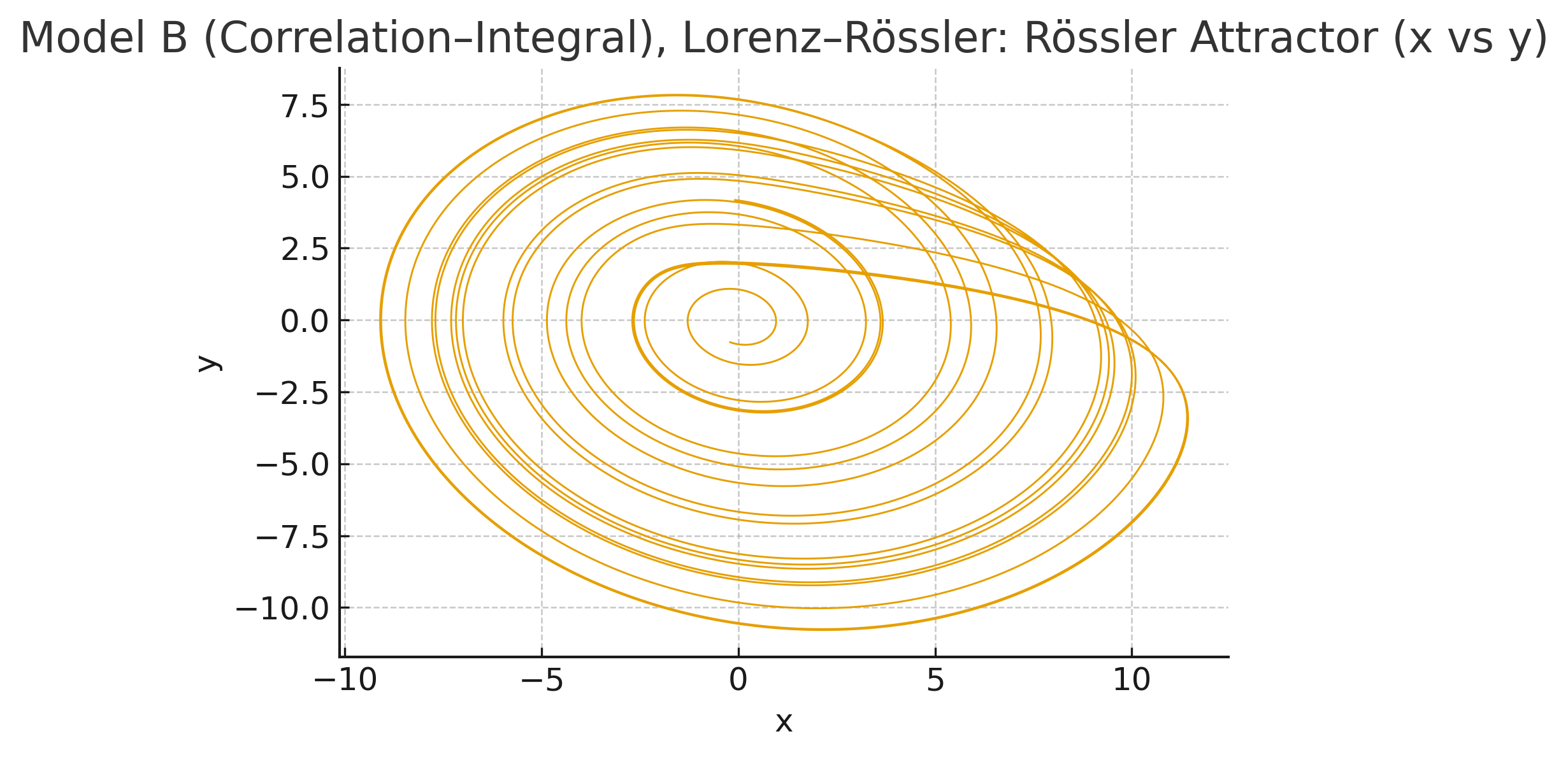}
  \caption{Rössler attractor ($x$ vs $y$)}
  \label{fig:4a}
\end{subfigure}\hfill
\begin{subfigure}{0.32\textwidth}
  \includegraphics[width=\linewidth]{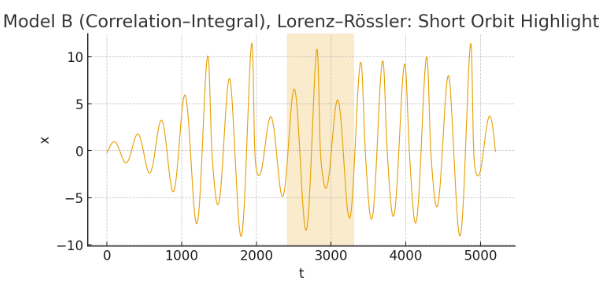}
  \caption{Short orbit highlight}
  \label{fig:4b}
\end{subfigure}\hfill
\begin{subfigure}{0.32\textwidth}
  \includegraphics[width=\linewidth]{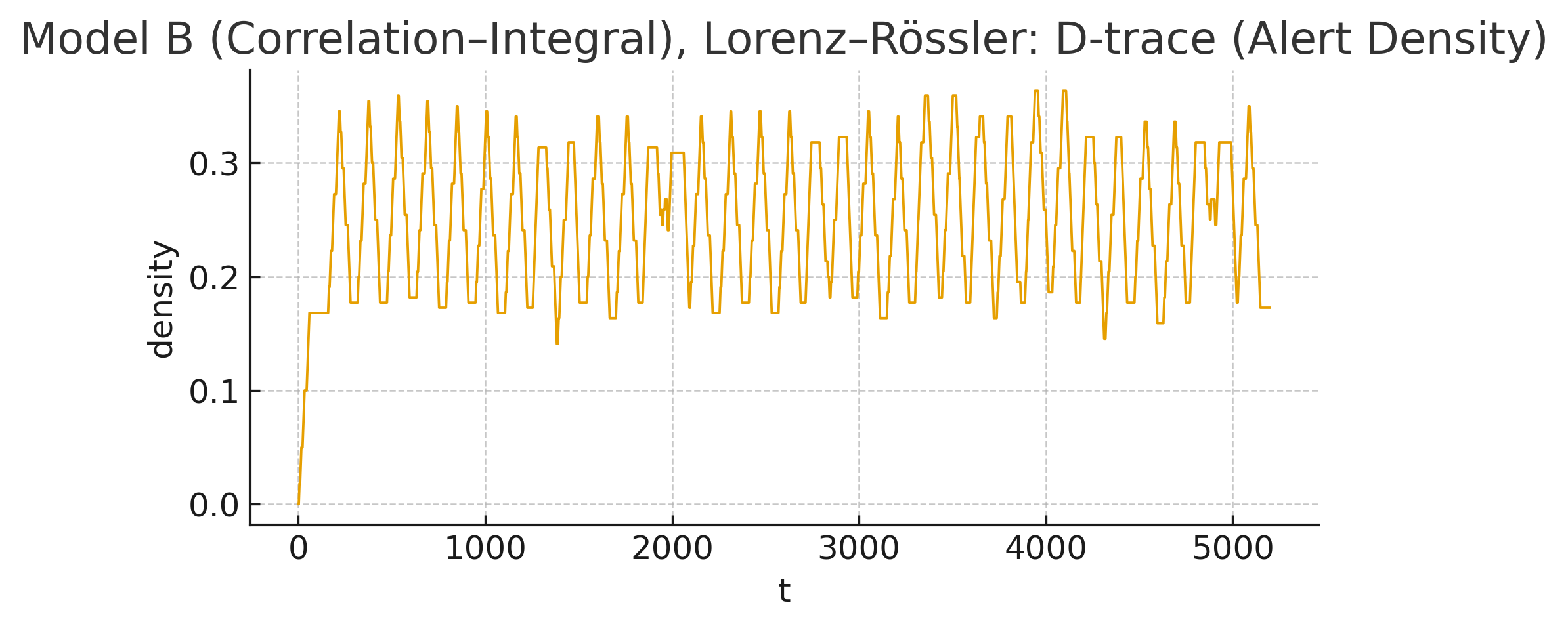}
  \caption{$D$-trace diagnostic}
  \label{fig:4c}
\end{subfigure}
\caption{Figure 4. Model~B (Correlation--Integral), Rössler attractor with Fibonacci diagnostics.}
\label{fig:4}
\end{figure}

\subsection{Supplementary Baseline Diagnostics}

For comparison, we computed rolling volatility estimates and standardized returns using fixed windows 
(50 and 200 steps). 
These conventional diagnostics captured long-run variance shifts but smoothed over short-lived bursts, 
limiting sensitivity to clustering. 
A direct comparison showed that Fibonacci diagnostics detect bursts across multiple scales 
where fixed windows failed (Suppl.~Fig.~S1a--c). 

\clearpage
\setcounter{figure}{0}
\renewcommand{\thefigure}{S\arabic{figure}}

\begin{figure}[H]
\centering
\begin{subfigure}{0.32\textwidth}
  \includegraphics[width=\linewidth]{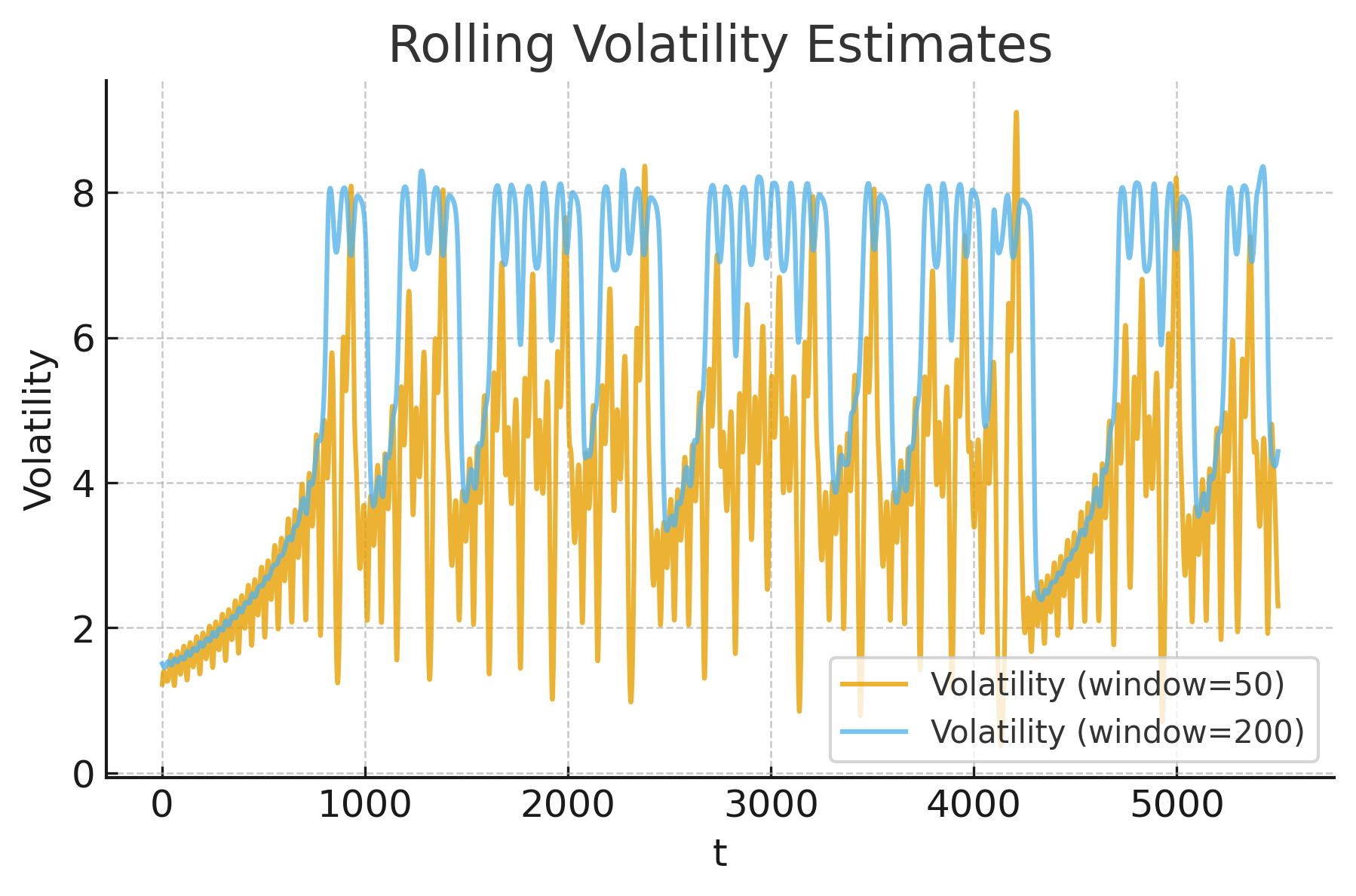}
  \caption{Rolling volatility (50, 200)}
  \label{fig:S1a}
\end{subfigure}\hfill
\begin{subfigure}{0.32\textwidth}
  \includegraphics[width=\linewidth]{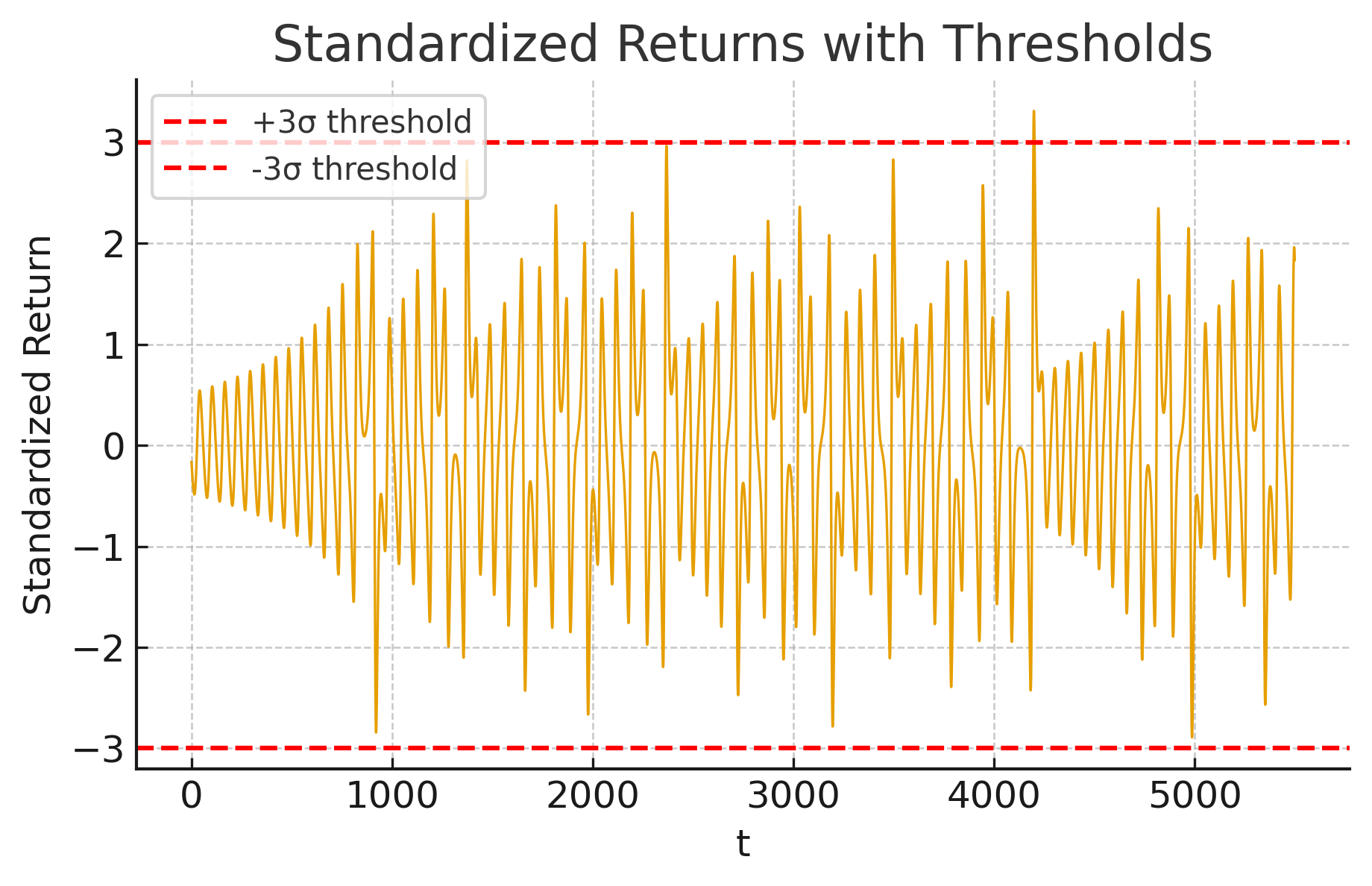}
  \caption{Standardized returns $\pm 3\sigma$}
  \label{fig:S1b}
\end{subfigure}\hfill
\begin{subfigure}{0.32\textwidth}
  \includegraphics[width=\linewidth]{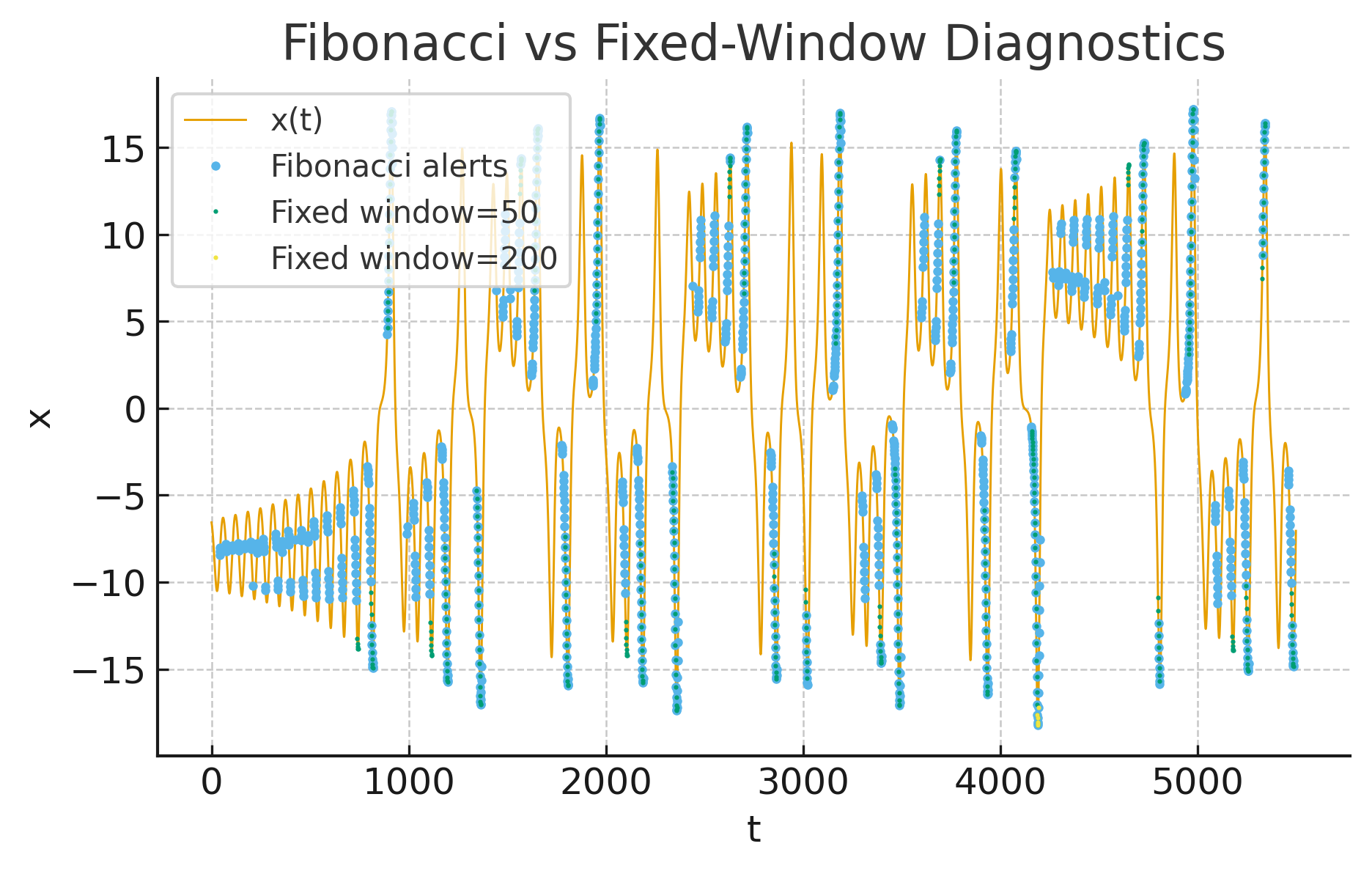}
  \caption{Fibonacci vs fixed-window alerts}
  \label{fig:S1c}
\end{subfigure}
\caption{Supplementary Figure S1. Baseline diagnostics.}
\label{fig:S1}
\end{figure}

\subsection{Combined Interpretation}

The Lorenz--Lorenz experiment demonstrates that the two models yield qualitatively different posteriors 
even on the same system: Model~A anchors inference in attractor geometry, while Model~B foregrounds 
rare-event clustering. 
The Lorenz--Rössler experiment shows that the Correlation--Integral framework generalizes beyond Lorenz, 
capturing volatility structure in the Rössler attractor. 
Supplementary comparisons with fixed-window diagnostics further highlight the novelty of the Fibonacci approach. 
Together, these results establish a dual-model framework that integrates geometric stability 
with tail-sensitive volatility detection.This confirmed that the geometric approach anchored inference in baseline attractor structure.

\section{Discussion}

The dual-model analysis highlights how inference outcomes depend on whether geometry or recurrence 
is prioritized. Model~A (Poincaré--Mahalanobis) consistently anchored inference in the attractor’s 
geometric structure. By leveraging Poincaré sections and Mahalanobis discrepancy, it reliably 
recovered canonical parameter regimes and reproduced the expected attractor geometry. 
This makes Model~A a natural reference for baseline system stability. 

Model~B (Correlation--Integral with Fibonacci diagnostics) departed from this baseline by 
emphasizing recurrence density and burst detection. The correlation integral captured 
scaling structure, while Fibonacci-window diagnostics flagged volatility bursts at recursive 
timescales. The resulting posterior distributions favored parameter regimes that generated 
rare-event clustering, even when pointwise trajectory fidelity was reduced. 
This highlights the novelty of Model~B: it re-weights inference toward tail-sensitive statistics, 
providing a fundamentally different view of the same dynamical system. 

\section{Implications for Risk Analysis}

The two models provide complementary insights for systemic risk analysis. 
Model~A serves as a baseline tool, recovering stable geometric structure 
and canonical parameter values. This anchors analysis in expected system 
behavior and provides interpretability for conventional practice. 
Model~B, by contrast, highlights volatility bursts and rare-event regimes. 
The use of Fibonacci windows ensures that bursts are detected across 
multiple scales, capturing instabilities that fixed-window diagnostics miss. 
The $D$-trace offers a direct record of how well proposals reproduce burst 
structure, creating a volatility-sensitive diagnostic stream.

In practice, this dual-model perspective allows analysts to integrate two 
views: the stable attractor geometry (Model~A) and the volatility-sensitive 
burst profile (Model~B). For risk management, this is particularly valuable: 
one model grounds expectations, the other foregrounds low-probability but 
high-impact deviations. Together, they provide a richer basis for anticipating 
rare events and Black Swan dynamics.

\section{Conclusion}

We introduced a dual-model framework for chaotic inference and rare-event detection. 
Model~A (Poincaré--Mahalanobis) anchors inference in geometric structure, 
while Model~B (Correlation--Integral with Fibonacci diagnostics) emphasizes recurrence statistics 
and volatility clustering. Applied in tandem, the models provide complementary insights. 

The Lorenz--Lorenz experiments demonstrated that even within a single system, 
diagnostic weighting can fundamentally alter posterior inference, shifting emphasis 
from baseline stability to rare-event regimes. The Lorenz--Rössler experiments showed 
that the correlation-integral framework generalizes across distinct attractors, 
retaining sensitivity to volatility clustering despite different dynamical profiles. 

Taken together, these results establish that the Poincaré--Mahalanobis model provides 
a stable geometric anchor, while the Correlation--Integral model introduces a novel 
and robust pathway for rare-event detection. This dual perspective advances both 
methodological understanding and practical tools for systemic risk analysis. 
Future work will extend the framework to higher-dimensional attractors, 
optimize computational efficiency for ABC-MCMC, and apply the models to 
empirical domains such as financial markets, climate systems, and infrastructure resilience. 

\par
\section{Future Work}

Future work includes calibration to real financial data, extension to 
higher-dimensional attractors, and integration with agent-based models 
for systemic risk. In particular, calibration under fat-tailed priors 
will allow direct comparison between Model~A (Poincaré--Mahalanobis) 
and Model~B (Correlation--Integral) when exposed to empirical volatility 
clustering. Extending the dual-model framework to higher-dimensional 
attractors (e.g., coupled Lorenz--Rössler systems) would test robustness 
to additional nonlinearities, while agent-based integrations would 
connect attractor dynamics to systemic cascades in financial networks. 
These directions will help establish whether the dual approach can 
provide early-warning signals of rare, destabilizing events in practice.

\section{Appendix}

\section*{Supplementary Figures}

Conventional risk diagnostics often rely on rolling volatility estimates and standardized returns. 
These approaches provide a useful baseline, but they are limited by their reliance on fixed windows, 
which can obscure the timing and intensity of rare-event bursts. 
The following supplementary figures illustrate these baseline methods, 
against which our proposed Fibonacci-window diagnostics can be compared.

\begin{figure}[h]
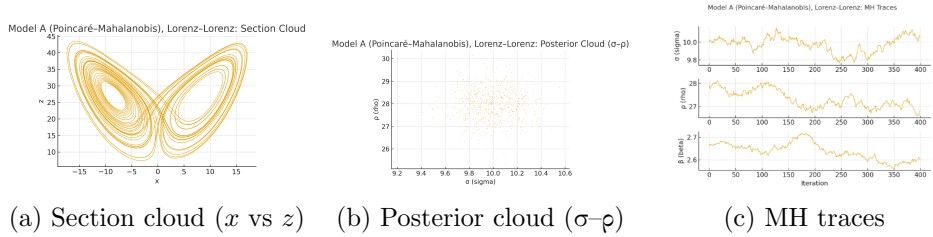

\centering
\begin{subfigure}{0.32\textwidth}
\includegraphics[width=\linewidth]{figA_section_cloud_LL.png}
\caption{Section cloud ($x$ vs $z$)}
\end{subfigure}
\hfill
\begin{subfigure}{0.32\textwidth}
\includegraphics[width=\linewidth]{figA_post_cloud_LL.png}
\caption{Posterior cloud (σ--ρ)}
\end{subfigure}
\hfill
\begin{subfigure}{0.32\textwidth}
\includegraphics[width=\linewidth]{figA_mh_traces_LL.png}
\caption{MH traces}
\end{subfigure}
\caption{Figure~1. Model~A (Poincaré--Mahalanobis), Lorenz--Lorenz. 
(a) Section cloud reproduces Lorenz butterfly geometry. 
(b) Posterior clouds concentrate around canonical values. 
(c) MH traces show stable mixing around true parameters.}
\end{figure}

\begin{figure}[h]
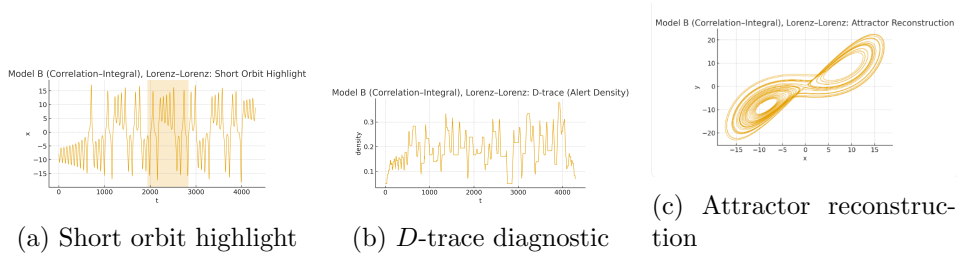

\centering
\begin{subfigure}{0.32\textwidth}
\includegraphics[width=\linewidth]{figB_short_orbit_LL.png}
\caption{Short orbit highlight}
\end{subfigure}
\hfill
\begin{subfigure}{0.32\textwidth}
\includegraphics[width=\linewidth]{figB_dtrace_LL.png}
\caption{$D$-trace diagnostic}
\end{subfigure}
\hfill
\begin{subfigure}{0.32\textwidth}
\includegraphics[width=\linewidth]{figB_attractor_LL.png}
\caption{Attractor reconstruction}
\end{subfigure}
\caption{Figure~2. Model~B (Correlation--Integral), Lorenz--Lorenz. 
(a) Short orbit segments flagged by Fibonacci windows ($21/34/55/89$). 
(b) $D$-trace records diagnostic distances, with dips marking accepted proposals. 
(c) Attractor reconstruction confirms Lorenz structure under diagnostic weighting.}
\end{figure}

\begin{figure}[h]
\centering
\includegraphics[width=0.6\textwidth]{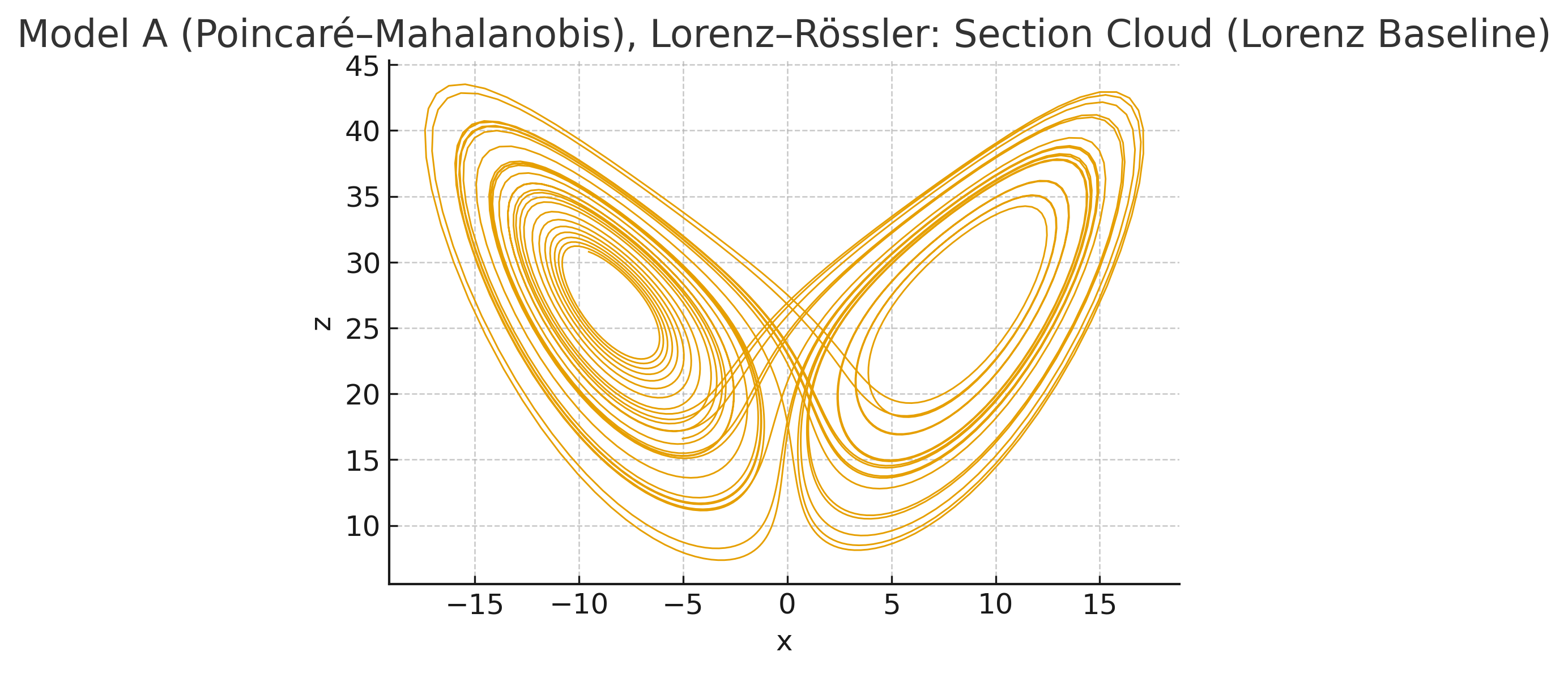}
\caption{Figure~3. Model~A (Poincaré--Mahalanobis), Lorenz--Rössler pairing. 
Section cloud of Lorenz attractor used as geometric baseline anchor.}
\end{figure}

\begin{figure}[h]
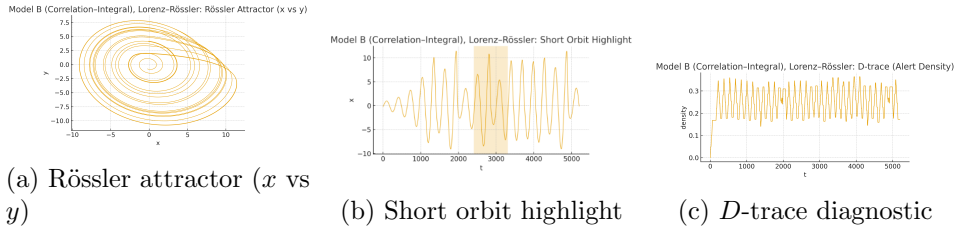

\centering
\begin{subfigure}{0.32\textwidth}
\includegraphics[width=\linewidth]{figB_attractor_LR.png}
\caption{Rössler attractor ($x$ vs $y$)}
\end{subfigure}
\hfill
\begin{subfigure}{0.32\textwidth}
\includegraphics[width=\linewidth]{figB_short_orbit_LR.png}
\caption{Short orbit highlight}
\end{subfigure}
\hfill
\begin{subfigure}{0.32\textwidth}
\includegraphics[width=\linewidth]{figB_dtrace_LR.png}
\caption{$D$-trace diagnostic}
\end{subfigure}
\caption{Figure~4. Model~B (Correlation--Integral), Rössler attractor. 
(a) Spiral geometry of the Rössler system. 
(b) Short orbit highlights volatility bursts identified by Fibonacci diagnostics. 
(c) $D$-trace records diagnostic mismatch, with dips marking accepted proposals.}
\end{figure}

\begin{figure}[h]
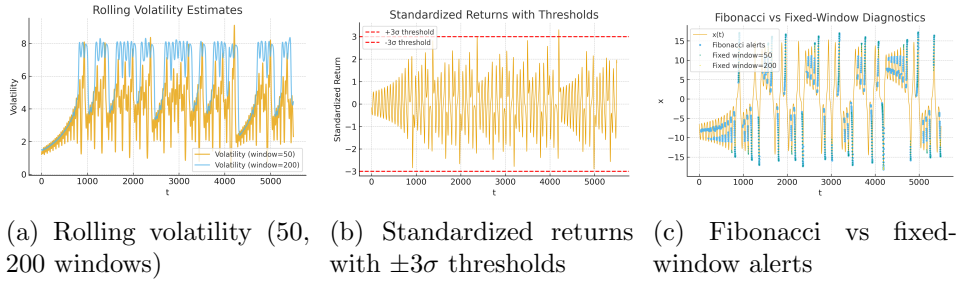

\centering
\begin{subfigure}{0.32\textwidth}
\includegraphics[width=\linewidth]{fig_roll_volatility.png}
\caption{Rolling volatility (50, 200 windows)}
\end{subfigure}
\hfill
\begin{subfigure}{0.32\textwidth}
\includegraphics[width=\linewidth]{fig_standardized_returns.png}
\caption{Standardized returns with $\pm 3\sigma$ thresholds}
\end{subfigure}
\hfill
\begin{subfigure}{0.32\textwidth}
\includegraphics[width=\linewidth]{fig_fib_vs_fixed.png}
\caption{Fibonacci vs fixed-window alerts}
\end{subfigure}
\caption{Supplementary Figure~S1. Baseline diagnostics. 
(a) Rolling volatility with fixed windows smooths over short bursts. 
(b) Standardized returns highlight extreme deviations but depend strongly on window size. 
(c) Fibonacci-window diagnostics capture burst clustering across scales, outperforming fixed-window methods.}
\end{figure}

\end{document}